\documentclass[twocolumn,prb,showpacs,floatfix]{revtex4}

\usepackage{graphicx}
\usepackage{bm}
\usepackage{color}
\usepackage{amsmath}
\usepackage{amsfonts}
\usepackage{amssymb}

\begin{document}

\preprint{APS/123-QED}

\title{Thermal and magnetic properties of a low-temperature antiferromagnet Ce$_4$Pt$_{12}$Sn$_{25}$}

\author{Nobuyuki Kurita$^{1}$} \author{Han-Oh  Lee$^{1,2}$} \author{Yoshi Tokiwa$^1$} \author{Corneliu F. Miclea$^1$}  
\author{Eric D. Bauer$^1$}  \author{Filip Ronning$^1$} \author{J. D. Thompson$^1$} \author{Zachary Fisk$^2$} \author{Pei-Chun Ho$^{3,4}$} \author{M. Brian Maple$^3$}
\author{Pinaki Sengupta$^{1,5}$} \author{Ilya Vekhter$^6$}\author{Roman Movshovich$^1$}
\affiliation{$^1$Los Alamos National Laboratory, Los Alamos, New Mexico 87545, USA}
\affiliation{$^2$University of California, Irvine, California 92697, USA} 
\affiliation{$^3$Department of Physics, University of California, San Diego, La Jolla, California 92093, USA }
\affiliation{$^4$Department of Physics, California State University, Fresno, Fresno, California 93740, USA}
\affiliation{$^5$School of Physical and Mathematical Sciences, Nanyang Technological University, 21 Nanyang Link, Singapore 637371} 
\affiliation{$^6$Department of Physics \& Astronomy, Louisiana State University, Baton Rouge, LA 70803, USA}

\date{\today}

\begin{abstract}
We report specific heat ($C$) and magnetization ($M$) of single crystalline Ce$_4$Pt$_{12}$Sn$_{25}$ at temperature down to $\sim$\,50\,mK and in fields up to 3\,T. 
$C/T$ exhibits a sharp anomaly at 180\,mK, with a large $\Delta C$/$T$\,$\sim$\,30\,J/mol\,K$^2$-Ce, which, together with the corresponding cusp-like magnetization anomaly, 
indicates an antiferromagnetic (AFM) ground state with a N\'eel temperature $T_\mathrm{N}$\,=\,180\,mK. 
Numerical calculations based on a Heisenberg model reproduce both zero-field
$C$ and $M$ data, thus placing Ce$_4$Pt$_{12}$Sn$_{25}$ in the weak exchange
coupling $J$\,$<$\,$J_\mathrm{c}$ limit of the Doniach diagram, with a very small Kondo scale $T_\mathrm{K}$\,$\ll$\,$T_\mathrm{N}$. 
Magnetic field suppresses the AFM state at $H^*$\,$\approx$\,0.7\,T, much more effectively than expected  from the
Heisenberg model, indicating additional effects possibly due to frustration or residual Kondo screening.


\end{abstract}

\maketitle

\section{Introduction}


Compounds with elements containing 4$f$\,- and 5$f$\,-\,electrons, such as Ce, Yb, and U, have been
subjects of intense research over the past several decades. Rich physics displayed within this class of
materials can be directly attributed to the interaction of $f$\,-\,electrons with conduction electrons,
which leads to, among other aspects, competing ground states. Heavy electron ground states are often
thought to be a consequence of the Kondo exchange interaction, $J$, between the $f$\,- and
conduction $c$\,-\,electrons\,~\cite{Kondo:PTP-64}.  At each site this interaction favors the
formation of the local $f$\,-\,$c$ Kondo singlet and leads to large effective masses of
 the charge carriers. The same $f$\,-\,$c$ exchange term also leads to the
Ruderman\,-\,Kittel\,-\,Kasuya\,-\,Yosida (RKKY)~\cite{Ruderman:PR-54,kasuya:ptp-56,yosida:PR-57} magnetic coupling
between the $f$\,-\,ions mediated by the polarization of the conduction electrons. The latter effect
generally favors long-range magnetic (often antiferromagnetic (AFM)) ground state of the $f$\,-\,ions.

Competition between the Kondo effect and RKKY interaction was originally addressed by Doniach in the
1D\,-\,chain model~\cite{doniach:physicaB-77,Doniach}, for different values of the AFM Kondo coupling.
At low $J$ the long\,-\,range AFM ground state of $f$\,-\,ions is stabilized. With increasing $J$ the
AFM ground state is suppressed, yielding to a Kondo screened state beyond a critical
value, $J_\mathrm{c}$. Pressure can be effective in changing on\,-\,cite $J$ by increasing the overlap between the
{\it f} and {\it c} states. A number of compounds were discovered that are located in the vicinity of
the $T$\,=\,$0$ phase transition between the heavy fermion (HF) and AFM ground states, often manifested as a Quantum Critical
Point (QCP). Non\,-\,Fermi\,-\,liquid (NFL) behavior displayed by compounds in the vicinity of QCPs, such as the
divergent Sommerfeld coefficient, $C/T$\,$\propto$\,$-$\,ln$T$ behavior found in
CeCu$_{5.9}$Au$_{0.1}$~\cite{lohneysen:JPCM-96}, YbRh$_2$Si$_2$~\cite{gegenwart:prl-02}, and
CeCoIn$_5$~\cite{bianchi:prl-03b,paglione:prl-03}, linear, in temperature, resistivity, and other
anomalous properties, provides additional impetus to research on competition between HF
physics and magnetism. The NFL behavior in $d$\,- and $f$\,-\,electron systems is reviewed, for
example, in Refs.~\onlinecite{stewart:rmp-01,stewart:rmp-06}.

There have been a number of theoretical attempts to model such systems using the periodic Anderson
model~\cite{lee:commSSP-86} and Kondo lattice model~\cite{sinjukow:prb-02}. A mean field model of the
Kondo lattice including the nearest\,-\,neighbor magnetic interaction~\cite{iglesias:prb-97} yields a
reduction in the Kondo temperature, $T_\mathrm{K}$, at $J$\,$>$\,$J_\mathrm{c}$ compared to its value for a single
impurity. This result emphasizes the importance of the RKKY coupling even when the system is in the
Kondo (HF) ground state. On the other hand, for $J$\,$<$\,$J_\mathrm{c}$, on the AFM side of the
QCP in the Doniach analysis, a mean field renormalization group approach~\cite{rappoport:JPA-01}
provided a good agreement between theory and the experimental results~\cite{cornelius:prb-94} on the
pressure-tuned AFM transition temperatures in a series of Ce$T_2$Si$_2$ ($T$\,=\,Ru, Rh, and Pd)
compounds. However, the interplay between the Kondo and RKKY interactions in this regime is not yet
fully understood~\cite{Yang}, and additional work, both theoretical and experimental, is needed. In this paper we
provide an example of a compound which orders magnetically at low temperature, in the regime where
previously studied compounds display dominant Kondo screening.

An alternative to the pressure route of tuning the relative strength of the Kondo screening and AFM
coupling is via synthesis, i.e. crystallographic structure of the compounds. For a specific value of the
on\,-\,site $f$\,-\,$c$ exchange constant $J$ one expects the RKKY interaction, which decays as
a power law in $k_Fr$, where $k_F$ is the Fermi momentum and $r$ is the distance between the magnetic
ions, to become weaker the further the $f$\,-\,atoms are separated from one another. As a result, the
Kondo screening should commonly win over magnetic ordering in such dilute systems. It would therefore be
particularly interesting to explore the $f$\,-\,electron bearing compounds with large $f$\,-\,$f$ nearest
neighbor distances. Several stoichiometric dilute $f$\,-\,electron compounds have been useful in
this regard. Ce\,-\,based filled skutterudites Ce$T_4X_{12}$ ($T$\,$=$\,Fe, Ru, Os; $X$\,$=$\,P, As, Sb,
$d_\mathrm{Ce-Ce}$\,$\sim$7\,$\mathrm{\AA}$) form one of these families, with most members semiconducting
with the gap size that correlates with the lattice constant\,\cite{Sugawara:prb-05}. Another example is
a family of Yb-based HF compounds Yb$A_2$Zn$_{20}$ ($A$\,$=$\,Fe, Co, Ru, Rh, Os, Ir,
$d_\mathrm{Yb-Yb}$\,$\sim$6\,$\mathrm{\AA}$)~\cite{torikachvili:pnas-07,jia:prb-08,YbCo2Zn20-saiga:jpsj-08},
where the Kondo physics appears to dominate the RKKY interaction, in accord with the simple argument
above.

The approach of synthesizing compounds with large $f$\,-\,$f$ nearest neighbor distance also turned out to
be useful in the studies of electronic correlations in uranium compounds. Large U\,-\,U distance
$d_\mathrm{U-U}$\,$\sim$6\,$\mathrm{\AA}$ in U$M_2$Zn$_{20}$ ($M$\,$=$\,Co, Rh) family of materials reduces the overlap
between the 5$f$\,-\,electron wave functions, allowing for observation of sharp crystal electric fields,
a situation very rare in U compounds\,~\cite{bauer:UM2Zn20-prb-08}. At the same time, this U\,-\,U separation
resulted in a good description of the system within a Kondo limit of the Anderson model, i.e.
U$M_2$Zn$_{20}$ compounds are found to be well on the Kondo side of the QCP, similar to the
Yb$T_2$Zn$_{20}$ case above, and again in accord with expectations above.

Ce$_4$Pt$_{12}$Sn$_{25}$ presents another example of a metallic 4$f$\,-\,dilute Ce\,-\,based
system~\cite{Chafik}, with Ce\,-\,Ce interatomic distance $d_\mathrm{Ce-Ce}$\,=\,6.14\,$\mathrm{\AA}$. 
It therefore appears to be a good candidate to continue exploration of correlated electron physics in $f$\,-\,electron systems, and in particular the competition between the RKKY interaction and Kondo screening.
Ce$_4$Pt$_{12}$Sn$_{25}$ is a cubic compound lacking 4\,-\,fold point symmetry with three inequivalent Sn sites.
Recently, we succeeded in growing single crystals of Ce$_4$Pt$_{12}$Sn$_{25}$ with relatively large
physical dimensions (up to $\sim$\,5$\times$5$\times$5\,mm$^3$)\,\cite{HanOh}. Previous specific heat,
resistivity, and ac\,-\,susceptibility  measurements uncovered a phase transition at 0.18\,K.~\cite{HanOh}
However, the origin of the transition was not identified. In this paper, we report the low-temperature
specific heat ($C$) and magnetization ($M$) measurements in magnetic field. From the cusp and the field dependence
in $M$($T$), which corresponds to the specific heat anomaly, we conclude that Ce$_4$Pt$_{12}$Sn$_{25}$
is an antiferromagnet with $T_\mathrm{N}$\,=\,0.18\,K in zero field. Evolution of  $C/T$ and $M$ with
magnetic field indicates disappearance of AFM ordering and subsequent splitting of an $f$\,-\,electron
ground state doublet. We find that the signatures of the magnetic transition are reasonably close to
those described by a spin\,-\,1/2 Heisenberg model, albeit with notable differences, see below. This
suggests that even for the large distance between the neighboring $4f$\,-\,ions the RKKY interaction plays a
very significant, or even dominant, role.

\section{Experimental Details}
Single crystals of Ce$_4$Pt$_{12}$Sn$_{25}$ were grown by Sn self-flux method. The details of the sample
growth and the physical properties are described in Ref.\,\onlinecite{HanOh}. Specific heat was measured
in a SHE dilution refrigerator with 9\,T superconducting magnet by means of a quasi-adiabatic heat pulse
method with a RuO$_2$ thermometer. The low-temperature high-resolution DC magnetization measurements
were performed in a commercial Oxford Kelvinox dilution refrigerator with 12\,-\,14\,T superconducting
magnet. We used a capacitive Faraday magnetometer cell with applied field gradient of 10\,T/m. The
principle of the magnetization measurement is described in Ref.\,\onlinecite{Sakakibara}. By comparing
the data from a commercial Magnetic Property Measurement System (MPMS; Quantum design) up to 7\,T and
down to 2\,K with that obtained with capacitive method in the same field and temperature range, we are
able to determine the absolute value of the magnetization. We used three samples from different batches
labeled as samples \#1 (7.18\,mg), \#2 (2.58\,mg) and \#3 (3.70\,mg) in the figures in this paper for
specific heat measurements. We used a sample weighing 4.63\,mg for magnetization measurements.

\begin{figure}
\begin{center}
\includegraphics[width=0.95\linewidth]{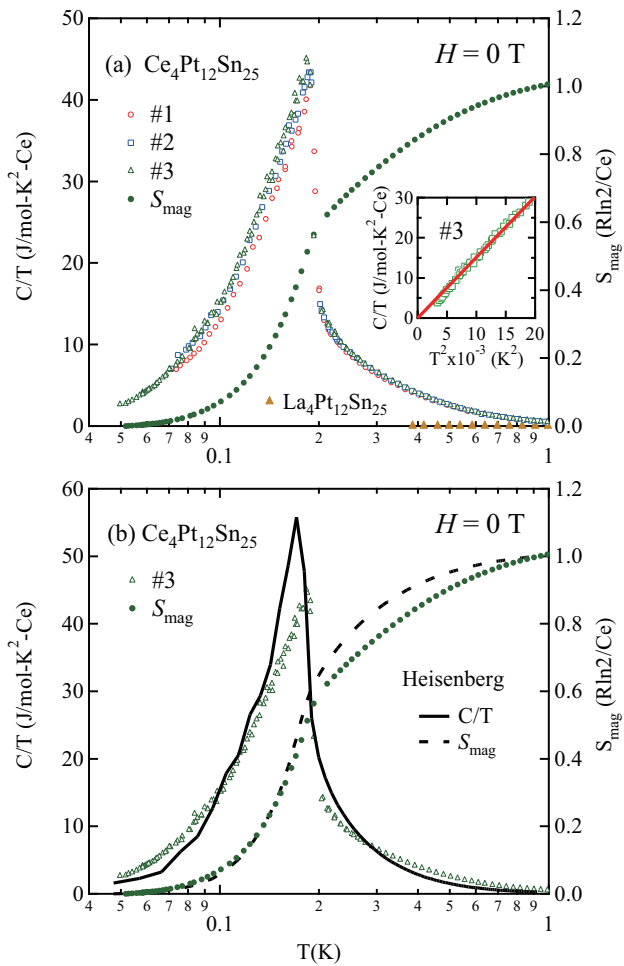}
\end{center}
\caption{(Color online) (a) Temperature dependence of $C$/$T$ of Ce$_4$Pt$_{12}$Sn$_{25}$ for samples
\#1, \#2 and \#3, and of the La-analog La$_4$Pt$_{12}$Sn$_{25}$ in zero field. Solid dots represent the
temperature dependence of magnetic entropy $S_\mathrm{mag}$ in units of $R$ln\,2. Inset: $C$/$T$ vs
$T^2$ at low temperature. The solid line is a linear least-squared fit to the data. (b) $\triangle$ -
$C$/$T$ and $\bullet$ - $S$ of Ce$_4$Pt$_{12}$Sn$_{25}$ (sample \#3). Solid and dashed lines represent
specific heat and entropy, respectively, from the Heisenberg model calculations described in the text. }
\label{fig1}
\end{figure}

\begin{figure}
\begin{center}
\includegraphics[width=0.95\linewidth]{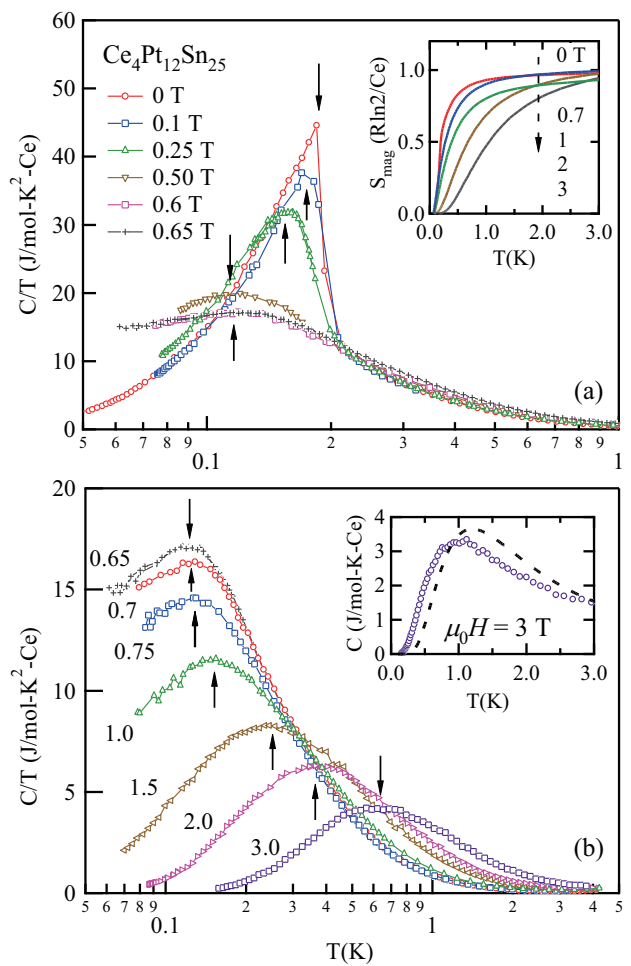}
\end{center}
\caption{(Color online) Temperature dependence of $C$/$T$ of Ce$_4$Pt$_{12}$Sn$_{25}$ in the magnetic
fields, (a) 0-0.65\,T, (b) 0.6-3\,T. The arrows indicate the temperature where $C/T$ has maximum. The
inset (a): $S_\mathrm{mag}$($T$) in several fields up to 3\,T. The inset (b): $C(T)$ at $B$\,$=$\,3\,T.
The dashed curve is obtained based on the CEF calculation defined in the text.} \label{fig2}
\end{figure}

\section{Results and Discussions}
\subsection{Specific heat}
Figure~\ref{fig1} shows the temperature dependence of $C$/$T$ of Ce$_4$Pt$_{12}$Sn$_{25}$
(50\,mK\,-\,3\,K) for samples \#1, \#2 and \#3, and that of a non-magnetic analogue
La$_4$Pt$_{12}$Sn$_{25}$ (0.4\,K\,-\,3\,K) in zero field. $C$/$T$ of La$_4$Pt$_{12}$Sn$_{25}$ is
negligible compared to that of Ce$_4$Pt$_{12}$Sn$_{25}$ in the investigated temperature range.
Therefore, the obtained specific heat of Ce$_4$Pt$_{12}$Sn$_{25}$ can be regarded as purely of magnetic
origin $C_\mathrm{mag}$. With temperature decreasing from 3 K, $C$/$T$ monotonically increases down to
0.18 K, at which point it exhibits a sharp anomaly with the magnitude of the jump
$\Delta C$/$T$\,$\sim$\,30\,J/mol\,K$^2$-Ce. From magnetization measurements we identify this anomaly
as due to antiferromagnetic (AFM) ordering, as discussed below. In addition, low-temperature
$T^2$-variation of $C$/$T$, shown in the inset, is consistent with the spin-wave contribution in the AFM
state. Solid circles represents the temperature dependence of the magnetic entropy $S_\mathrm{mag}(T)$
of Ce$_4$Pt$_{12}$Sn$_{25}$:
\begin{eqnarray}
S_\mathrm{mag}(T)=\int_0^T {{\frac{C_\mathrm{mag}}{T}}\,\mathrm{d}T}
\end{eqnarray}

Entropy gain is $\approx$ 0.5\,$R$ln\,2 at $T_\mathrm{N}$ and reaches the value of 1.0\,$R$ln\,2 at
3\,K. This indicates that the crystal electric field (CEF) ground state of Ce$^{3+}$ in
Ce$_4$Pt$_{12}$Sn$_{25}$ is a $\Gamma_7$ doublet. One remarkable feature of the specific heat of
Ce$_4$Pt$_{12}$Sn$_{25}$ is the long tail above $T_\mathrm{N}$. As we show below, a substantial part of
this tail is due to quantum fluctuations of $f$\,-\,electron spins above $T_\mathrm{N}$. However, from our analysis
it is likely that other processes, such as residual Kondo screening, or frustration of spin-spin
interactions play some role. Similar $C/T$ behavior was reported in the structurally frustrated system
Yb$_2$Pt$_2$Pb~\cite{Kim:prb-08}, for example.

To elucidate the origin of this behavior we performed Quantum Monte Carlo simulations of the spin\,-\,1/2
three-dimensional Heisenberg model
in zero and applied magnetic field. The exchange interaction was fixed to have the AFM ordering
temperature of $T_\mathrm{N}$\,=\,0.18\,K for $H$\,=\,0, while the $g$\,-\,factor of the spins was determined from the
saturation magnetization at high fields (see below). The results for zero field specific heat and
entropy are displayed in Fig.~\ref{fig1} as solid and dashed curves, respectively. The agreement between
the model calculations and experimental data is rather good. In particular, the behavior of the entropy
below $T_\mathrm{N}$ is very similar, and the entropy at $T_\mathrm{N}$ for model calculations is about 0.55\,$R$\,ln\,2,
about 10\% larger than the experimental value. The most notable difference is that experimental entropy
is lower than the calculated values above the transition, indicating that some degrees of freedom remain
``locked''. This discrepancy is also seen from the somewhat longer tail of the measured specific heat in
 Fig.~\ref{fig1}(b) relative to the Heisenberg model.
Such contribution may still be due to remnant Kondo physics. Recall that the distance between Ce ions is
large, and therefore RKKY exchange, and correspondingly the critical $J_\mathrm{c}$, is reduced. In the framework
of Doniach phase diagram, the only situation when RKKY interaction can still dominate the Kondo coupling
is when the bare exchange $J$ is very small. Ce$_4$Pt$_{12}$Sn$_{25}$ is therefore in the low $J$\,$<$\,$J_\mathrm{c}$ limit of the Doniach phase diagram, and the system orders magnetically below $T_\mathrm{N}$\,=\,0.18\,K. Note that
Ce\,-\,based skutterudite compounds and Yb$A_2$Zn$_{20}$, whose Ce\,-\,Ce and Yb\,-\,Yb distances are similar to
that in Ce$_4$Pt$_{12}$Sn$_{25}$, do not show magnetic ordering, and are in the $J$\,$>$\,$J_\mathrm{c}$ Kondo limit.
It should also be noted that there is no significant sample dependence for different samples \#1, \#2,
and \#3, with respect to the ordering temperature, the magnitude of the anomaly in $C$/$T$, and its
width.

Figure~\ref{fig2} shows $C/T$ for magnetic field (a) 0\,$\leq$\,$B$\,$\leq$\,0.65\,T and (b)
0.65\,$\leq$\,$B$\,$\leq$\,3.0\,T. As shown in Fig.~\ref{fig2}(a), the steep jump in $C/T$ associated
with AFM ordering is quickly suppressed, and $T_\mathrm{max}$, where $C/T$ has its maximum value,
gradually shifts to lower temperature, as indicated by arrows. On the other hand, Fig.~\ref{fig2}(b)
shows that with further increase of magnetic fields, $T_\mathrm{max}$ shifts to higher temperature,
while the peak height is continuously suppressed. This behavior can be understood as follows: In the low
field region the reduction of $T_\mathrm{max}$ is ascribed to the
suppression of the AFM ordering with field  ($T_\mathrm{max}$\,$\approx$\,$T_\mathrm{N}$). Around 0.7\,T, the feature associated with the AFM order is
suppressed entirely. Further increase of magnetic field increases Zeeman splitting of the crystal
electric field (CEF) doublet ground state (see below), leading to a Schottky anomaly in specific heat
with rising $T_\mathrm{max}$. Similar behavior has been observed in a number of other Ce compounds, e.g.
CeCu$_{5.9}$Au$_{0.1}$~\cite{Zeeman_CeCuAu}, CeCu$_2$Si$_2$~\cite{Zeeman_CeCu2Si2}, and (La,\,Ce)Al$_2$\,~\cite{Zeeman_CeCu2Si2,Zeeman_CeLaAl2}. As seen in the inset of Fig.~\ref{fig2}(a),
$S_\mathrm{mag}$ reaches 1.0$R$\,ln\,2 independently of applied fields up to 3\,T, although the ground
state and the shape of the specific heat anomaly both vary strongly with applied field, reflecting a
two\,-\,fold degeneracy of the CEF doublet.

At high magnetic fields,  far exceeding the Heisenberg exchange coupling, we can obtain the approximate
behavior of Ce$_4$Pt$_{12}$Sn$_{25}$ based on the CEF level scheme. To avoid complication in
parameterizing, we assume that a cubic point symmetry of Ce ions in Ce$_4$Pt$_{12}$Sn$_{25}$ is
$O_\mathrm{h}$ and the CEF Hamiltonian $H_\mathrm{CEF}$ can be reduced to the following
formula\,\cite{CEFmodification}:
\begin{eqnarray}
H_\mathrm{CEF}=\sum_{m,n}B_n^mO_n^m=B_4^0(O_4^0+5O_4^4) \label{CEF},
\end{eqnarray}
where $B_n^m$ and $O_n^m$ are the CEF parameters and the Stevens operators\,\cite{Stevens,Hutchings},
respectively. The total Hamiltonian $H$ and specific heat $C$ of CEF levels in external magnetic fields
is then given by the following expression:

\begin{eqnarray}
H=H_\mathrm{CEF}-g_J\mu_B\mbox{\boldmath $J$}\cdot\mbox{\boldmath $H$} \label{Hamiltonian}\\
C={\frac{\partial}{\partial{T}}}\frac{1}{Z}\sum_{n}E_n\mathrm{e}^{-\beta E_n} \label{HC}
\end{eqnarray}

Here, $g_J$ is the Lande $g$\,-\,factor, $E_\mathrm{n}$ and $|n\rangle$ are the $n$th eigenvalue and
eigenfunction, respectively, $Z$\,$=$\,$\sum_n e^{-\beta E_n}$, $\beta$\,=\,1/$k_\mathrm{B}T$ and
$k_\mathrm{B}$ is the Boltzmann constant. We assume that the ground state is a $\Gamma_7$ doublet.
$B_4^0$\,=0.5\,K then corresponds to the CEF splitting between $\Gamma_7$ and the first excited state
$\Delta$\,$\approx$\,200\,K, which can be inferred from $S_\mathrm{mag}$($T$) and resistivity
data\cite{HanOh}, respectively. As shown in the inset of Fig.~\ref{fig2}(b), the calculated specific
heat anomaly at 3\,T (dotted curve) closely reproduces experimental results, supporting the CEF scheme
suggested above.

\begin{figure}
\begin{center}
\includegraphics[width=0.95\linewidth]{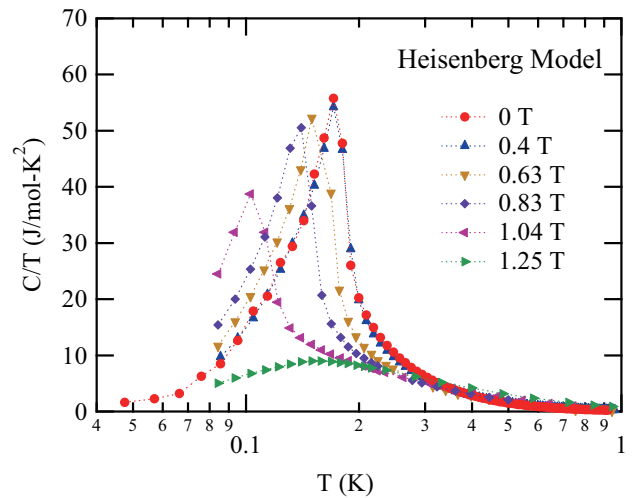}
\end{center}
\caption{Sommerfeld coefficient $\gamma$\,=\,$C/T$ calculated in the nearest neighbor Heisenberg model as
function of $T$ at low and intermediate fields . } \label{fig3}
\end{figure}

Quick suppression of the AFM specific heat anomaly in the low field regime, displayed in
Fig.~\ref{fig2}(a), and particularly the broadening of the anomaly with magnetic field, makes precise
identification of $T_\mathrm{N}(H)$ difficult. Encouraged by the success of our Heisenberg model calculations in
zero field, we performed calculations for this model in finite field, with the goal to aid in
identification of $T_\mathrm{N}$ in experimental data. Figure~\ref{fig3} displays the results of
these calculations. There are striking differences between the model and experimental data. 
In the Heisenberg model the specific heat anomaly at the AFM phase transition remains sharp, and persists to high fields over 1 Tesla, 
while the experimentally observed anomaly is washed out already by 0.5 T. 
There is a number of possible reasons for this discrepancy. A Heisenberg model is often used to describe
insulating compounds, with coupling $J$ independent of the magnetic field. In our case the magnetic
coupling is mediated by the conduction electrons, with a potential for a field-dependent magnetic
coupling. The effect of the next nearest neighbor interaction, with accompanying effects of possible
frustration, is also neglected in the present calculations. Among other complications is a possible
proximity to a HF (Kondo) ground state. Our results point to a number of fruitful future
theoretical inquiries, such as field dependent RKKY interaction, or including frustration within the
Heisenberg model due to next nearest neighbor interactions, and its response to magnetic field.

\begin{figure}
\begin{center}
\includegraphics[width=0.95\linewidth]{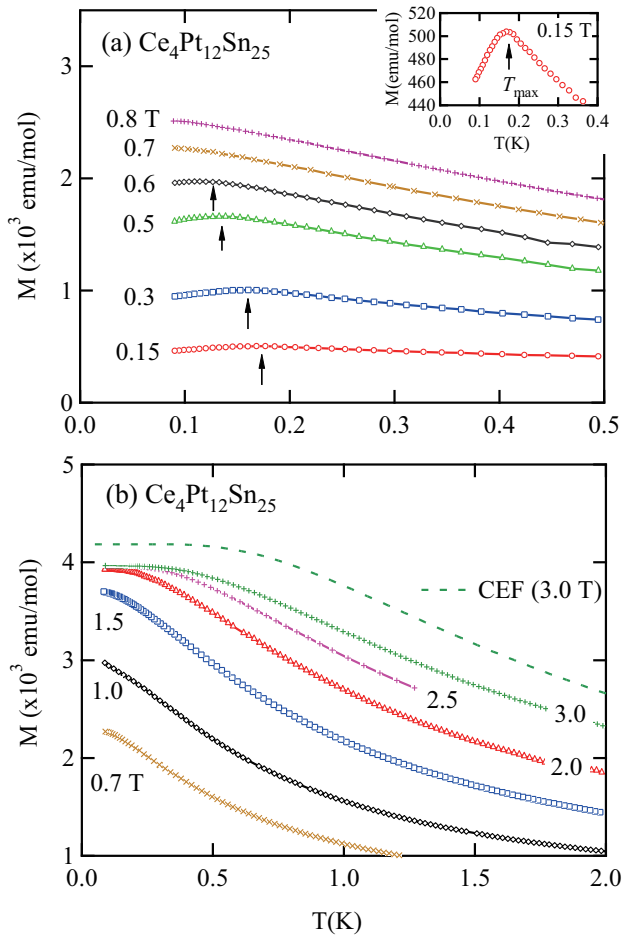}
\end{center}
\caption{(Color online) Temperature dependence of magnetization $M$($T$) of Ce$_4$Pt$_{12}$Sn$_{25}$ in
several fields (a) 0,-\,0.8\,T, (b) 0.7\,-\,3\,T. The arrows in (a) indicate $T_\mathrm{max}$ where
$M$($T$) exhibits maximum as displayed in the inset of (a).  The dashed curve in (b) represents a
calculated result at 3\,T based on the CEF effect (see text).} \label{fig4}
\end{figure}

\subsection{Magnetization}

Figure~\ref{fig4} shows temperature dependence of the magnetization $M$($T$) for $T$\,$\ll$\,2\,K in
fields (a) up to 0.8\,T and (b) between 0.7 and 3.0\,T. The low-temperature data connect well to the
higher temperature data obtained using a MPMS (not shown).  The cusp in $M$ vs. $T$ in the low field
region identified by arrows in Fig.~\ref{fig4}(a) reflects the AFM ordering phase transition. As an
example, the inset of Fig.~\ref{fig4}(a), shows $M$($T$) at 0.15\,T, on expanded scale, which exhibits a
clear cusp around $T_\mathrm{N}$\,=\,0.18\,K. It shifts to lower temperatures with field, in accord with the field
evolution of the specific heat anomaly. Together, specific heat and magnetization measurements prove
that Ce$_4$Pt$_{12}$Sn$_{25}$ undergoes an AFM phase transition at $T_\mathrm{N}$=0.18\,K in zero field.
With magnetic field increasing above 0.6\,T,  the cusp feature disappears, and the magnetization is
monotonic, suggesting that AFM order is suppressed in the vicinity of this field. Above 2.0\,T, the
magnetization develops a plateau at low temperature, as seen in Fig.~\ref{fig4}(b). The plateau is
related to the field splitting of the CEF ground state doublet. In fact, calculations based on the CEF
scheme assumed above also show similar behavior, as well as good agreement in absolute value, as
indicated by a dashed curve in the inset of Fig.~\ref{fig4}(b) obtained for 3\,T using the following
formula:

\begin{eqnarray}
\chi=\frac{N(g_J\mu_\mathrm{B})^2}{Z}[\sum_{m\neq n}|\langle m|J_z|n\rangle|^2
\frac{1-\mathrm{e}^{E_ m -E_n}}{E_m -E_n}\mathrm{e}^{-\beta
E_n}\\ \nonumber
+\sum_{n}|\langle n|J_z|n\rangle|^2 \beta\mathrm{e}^{-\beta E_n}]
\end{eqnarray}

\begin{figure}
\begin{center}
\includegraphics[width=0.95\linewidth]{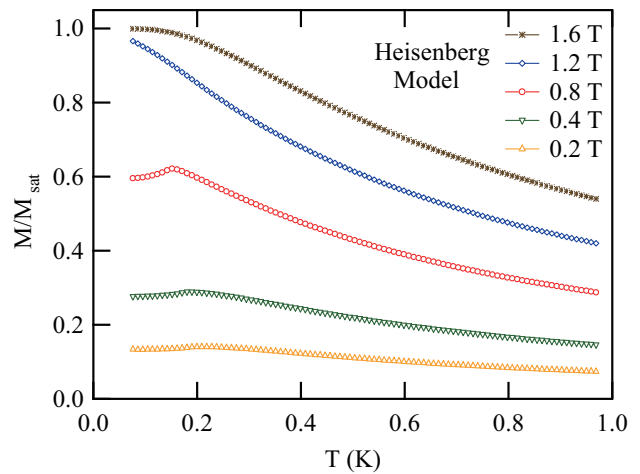}
\end{center}
\caption{Magnetization as a function of temperature in the Heisenberg model at different applied magnetic fields.}
\label{fig5}
\end{figure}

The results of the Heisenberg model calculations of magnetization are displayed in Fig.~\ref{fig5}. 
Similar to specific heat, the AFM transition remains very sharp and persists to higher
fields within the model calculations compared to the experimental data.


\begin{figure}
\begin{center}
\includegraphics[width=0.95\linewidth]{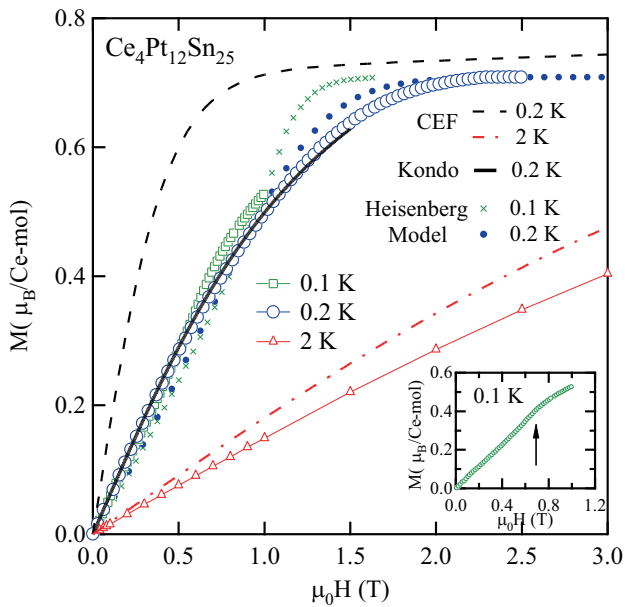}
\end{center}
\caption{(Color online) $M$ vs. $H$ curves of Ce$_4$Pt$_{12}$Sn$_{25}$ at several temperatures of 0.1,
0.2 and 2\,K. The dashed and dash-dotted curves are obtained from the calculation at 0.2\,K and 2 K,
respectively, based on a CEF scheme. Solid curve is a result of the Kondo screening model described in
the text. Dotted curve and crosses are numerical  results based on the Heisenberg model at 0.2 K and 0.1
K, respectively. The inset shows the magnetization curve at 0.1\,K. The arrow indicates the field where
the data show anomaly due to antiferromagnetic ordering.} \label{fig6}
\end{figure}

Figure~\ref{fig6} shows isothermal magnetization curves $M(H)$ of Ce$_4$Pt$_{12}$Sn$_{25}$ at
temperature of 0.1\,K ($<$\,$T_\mathrm{N}$), 0.2\,K ($\approx$\,$T_\mathrm{N}$) and 2\,K
($\gg$\,$T_\mathrm{N}$). In contrast to $M(H)$ data at 2.0\,K, which increases slowly with field and
does not show complete saturation up to 7\,T, $M(H)$ at 0.2\,K rapidly increases with field and exhibits
a plateau above 2\,T. The saturated value of the magnetic moment $\mu_\mathrm{s}$ corresponds to about
0.71\,$\mu_\mathrm{B}$/Ce, which is much smaller than the value expected for a free Ce$^{3+}$ ion of
$g_{J}$$J$$\mu_\mathrm{B}$\,=\, 2.14\,$\mu_\mathrm{B}$ ($g_{J}$\,=\,6/7, $J$\,=\,5/2). This reduction,
as well as deviation of $\chi$ at low temperature, reflects CEF effects in Ce$_4$Pt$_{12}$Sn$_{25}$.
Using formulas ~(\ref{CEF}),~(\ref{Hamiltonian}), and~(\ref{magnetization}) below, we can calculate the
magnetization by using the value of $B_4^0$\,$=$\,0.5\,K, the same as the one employed in calculations
of the specific heat above.

\begin{eqnarray}
M=\frac{g_J\mu_\mathrm{B}}{Z}\sum_{n}\langle n|J_z|n\rangle \beta\mathrm{e}^{-\beta E_n}
\label{magnetization}
\end{eqnarray}

The magnetization calculated from this CEF model is also displayed in Fig.~\ref{magnetization}, for
$T$\,=\,0.2\,K (dashed line) and for $T$\,=\,2\,K (dash-dotted line). Agreement between the CEF model and
the data is good for $T$\,=\,2\,K\,$\gg$\,$J$ (similar to that in Fig.~\ref{fig4}(b)). 
As expected, there is a large discrepancy at $T$\,=\,0.2\,K between experimental results and the calculated field dependence based on
a single ion CEF model, so that only the saturated
magnetic moment $\mu_\mathrm{s}$\,=0.74\,$\mu_\mathrm{B}$/Ce 
 is comparable to the experimental value ($\mu_\mathrm{s}$\,=\,0.71\,$\mu_\mathrm{B}$/Ce). 
The best fit to the data at 0.2\,K below 0.4\,T using a Kondo screening model\,\cite{Kondo_calculation}
gives $T_\mathrm{K}$\,=\,1.2\,K. However, this value of $T_K$ is incompatible with specific heat results
and the analysis performed above, which indicate $T_\mathrm{K}$\,$\le$\,0.2\,K~\cite{HanOh}. At intermediate fields the dominant physics is in the $f-f$ interactions inherent in the Heisenberg model that we
used to describe specific heat results above. The magnetization calculated for $T$\,=\,0.1\,K and 0.2\,K is
also displayed in Fig.~\ref{magnetization}, and indeed reproduces the experimental data well. Small
discrepancies at higher fields (calculations overestimate magnetization) that are more pronounced at low
temperature, are most likely again due to a weak Kondo screening. It should be noted that at 0.1\,K,
below $T_\mathrm{N}$ in zero field, $M$($H$) shows a kink anomaly around 0.6\,T, indicated by an arrow
in the inset of Fig.~\ref{magnetization}. This kink corresponds to the critical field $H^\mathrm{*}$
where AFM order disappears, and is consistent with $C(T)$/$T$ and $M$($T$) in constant magnetic fields
described above.

\begin{figure}
\begin{center}
\includegraphics[width=0.95\linewidth]{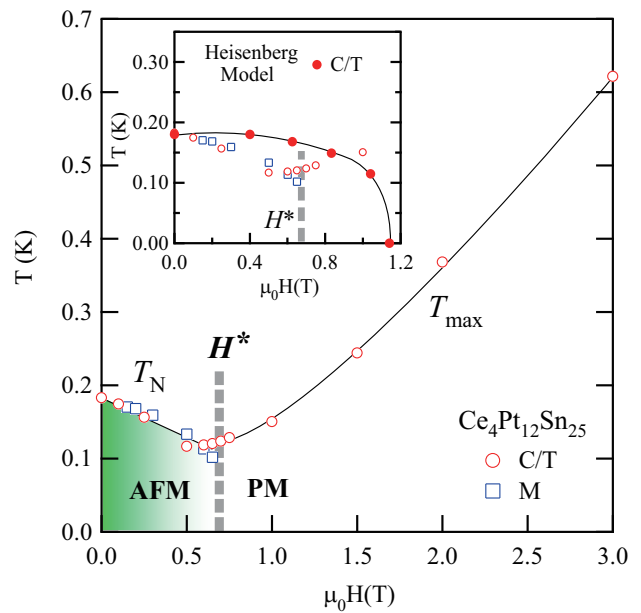}
\end{center}
\caption{(Color online) $T$\,-\,$H$ phase diagram of Ce$_4$Pt$_{12}$Sn$_{25}$ deduced from specific heat
 and magnetization measurements. Circle and square symbols correspond to $T_\mathrm{max}$ where $C(T)/T$
 and $M(T)$ exhibit maximum, respectively. The solid curve is the eye guide. The dashed line represents
 the critical magnetic field $H^\mathrm{*}$ which could be a boundary of the ground states between AFM
 (antiferromagnetic) and PM (paramagnetic) states. Inset shows the low field region, with results of the
 model calculations included.} \label{fig7}
\end{figure}

\subsection{$H$-$T$ phase diagram}

Figure~\ref{fig7} displays the temperature$-$field phase diagram of Ce$_4$Pt$_{12}$Sn$_{25}$ obtained from the
data described in previous sections, with circles and squares representing $T_\mathrm{max}$ obtained
from specific heat and magnetization measurements, respectively. $T_\mathrm{max}$
($\approx$\,$T_\mathrm{N}$) decreases as field increases up to about 0.6\,T, and $T_\mathrm{max}$
increases with further increase of magnetic field. $T_\mathrm{N}$ appears to vanish at a critical field
$H^\mathrm{*}$\,$\approx$\,0.7\,T indicated by the dashed line, where the ground state of
Ce$_4$Pt$_{12}$Sn$_{25}$ changes from AFM to paramagnetic. It is impossible to tell the exact manner in
which $T_\mathrm{N}$ goes to zero from the data presented here. 
As illustrated in the inset, the suppression of $T_\mathrm{N}$ with fields is much more effective
than expected from the Heisenberg model ($H^\mathrm{*}$\,$>$\,1\,T).
There is no divergence of Sommerfeld coefficient
$\gamma$ with $T$\,$\rightarrow$\,0, in contrast to common behavior in materials at a QCP. Perhaps this can
be explained by a small amount of entropy associated with magnetic fluctuations at high fields, as most
of it is released at a higher temperature Schottky anomaly. Low temperature spectroscopic
investigations, such as NMR, $\mathrm{\mu}$SR, or neutron scattering will be able to provide a definitive picture
of how AFM order is suppressed in Ce$_4$Pt$_{12}$Sn$_{25}$.

\section{Conclusion}
In conclusion, we have performed low-temperature field-dependent specific heat and magnetization
measurements to elucidate the ground state properties of Ce$_4$Pt$_{12}$Sn$_{25}$. Magnetization
measurements established that Ce$_4$Pt$_{12}$Sn$_{25}$ orders antiferromagnetically with a small
ordering temperature $T_\mathrm{N}$\,$=$\,0.18\,K in zero field, where $C(T)/T$ exhibits a huge jump of
$\sim$\,30\,J/mol\,K$^2$-Ce. The small value of the saturated magnetic moment of
0.71\,$\mu_\mathrm{B}$/Ce at 0.2\,K exhibited by magnetization as a function of magnetic field can be
ascribed to CEF effects, with a $\Gamma_7$ ground state, as inferred from magnetic entropy.
$T_\mathrm{N}$ is suppressed with an initial increase of field up to 0.6\,T, whereas $T_\mathrm{max}$ of
the maximum in $C(T)$ begins to move to higher temperature above 0.7\,T. This latter evolution is
ascribed to an electronic Schottky contribution from the Zeeman-split ground state $\Gamma_7$ doublet.
Therefore, it is likely that the ground state of Ce$_4$Pt$_{12}$Sn$_{25}$ changes above $\approx$\,0.6\,T
from AFM to paramagnetic. This picture is further supported by the magnetization data for both
temperature and magnetic field sweeps.

The model calculations, based on a CEF scheme with a $\Gamma_7$ ground state, reproduce rather well
experimental specific heat and magnetization data in high field. Zero field data are described very well
by the numerical calculation based on the Heisenberg model. Small deviation between experimental data
and numerical results may be due to frustration or Kondo screening with a low characteristic
temperature $T_\mathrm{K}$\,$\le$\,$T_\mathrm{N}$\,=\,0.18\,K. This indicates that electronic spins on Ce$^{3+}$ are not screened
substantially by the conduction electrons, and places Ce$_4$Pt$_{12}$Sn$_{25}$ in the $J$\,$\ll$\,$J_\mathrm{c}$ of the
Doniach phase diagram. The Kondo temperature $T_\mathrm{K}$ is expected to be exponentially small in this regime, and a
$J^2$ dependence of the RKKY interaction stabilizes the AFM ground state {\it in spite} of a large
distance between the Ce ions. Ce$_4$Pt$_{12}$Sn$_{25}$ therefore presents us with a {\it counter
example} to an expectation that dilute $f$\,-\,electron compounds will likely fall into the Kondo screened
regime.

Within the Heisenberg model calculations, the AFM transition is more robust to the magnetic field, compared
to experimental data in low field regime.  We raised a number of questions with regard to the possible
origins of this discrepancy, such as magnetic field dependence of the RKKY interaction, and frustrating
effects of the next nearest neighbor interactions, and point to a number of fruitful future theoretical
inquiries.


Further experiments will be needed to explore the details of the disappearance of the AFM state in
Ce$_4$Pt$_{12}$Sn$_{25}$, as well as the origin of the long temperature tail of the specific heat above
$T_\mathrm{N}$, and potential role of magnetic frustration. Experiments under pressure in particular can
help answer why $T_\mathrm{N}$ is so low in Ce$_4$Pt$_{12}$Sn$_{25}$ compared with other Ce compounds,
as well as search for pressure induced superconductivity, occasionally found in the vicinity of pressure
induced QCP at $P_\mathrm{c}$ ($T_\mathrm{N}$\,$\rightarrow$\,0).

\section*{ACKNOWLEDGMENTS}
We would like to thank Hironori Sakai for useful discussions. 
Work at Los Alamos National Laboratory was
performed under the auspices of the US Department of Energy. 
Research at UCSD supported by the US National Science Foundation under Grant No. DMR-0802478.
I. V. was supported in part by the US DOE
Grant No.~DE-FG02-08ER46492.

\end{document}